\documentclass[10pt]{mpi08} \usepackage{graphicx} 
\usepackage{cite,./mcite}  
\usepackage{amstext}
\usepackage{wrapfig}
\newcommand{\newc}{\newcommand}
\newc{\mathmode}[1]{\relax\ifmmode #1\else{$#1$}\fi}
\newc{\hwpp}{\textsf{Herwig++}}
\newc{\ptsum}{\mathmode{p_t^{\rm sum}}}
\newc{\ptmin}{\mathmode{p_t^{\rm min}}}
\newc{\vect}[1]{{\bf #1}}
\newc{\sjet}{\mathmode{\sigma_{\rm jet}}}
\newc{\sela}{\mathmode{\sigma_{\rm el}}}
\newc{\sinel}{\mathmode{\sigma_{\rm inel}}}
\newc{\sigmasoft}{\mathmode{\sigma^{\rm inc}_{\rm soft}}}
\newc{\sigmahard}{\mathmode{\sigma^{\rm inc}_{\rm hard}}}
\newc{\slope}{\mathmode{b_{\rm el}}}
\newc{\musq}{\mathmode{\mu^2}}
\newc{\ptmatch}{\mathmode{p_t^{\rm min}}}
\newc{\stot}{\mathmode{\sigma_{\rm tot}}}
\newc{\GeV}{\text{ GeV}}
\newc{\as}{\mathmode{\alpha_s}}
\newc{\di}[2]{{\rm d}^{#2}#1}
\renewcommand{\exp}[1]{\mathmode{\: e^{#1} \: }}
\newc{\smfrac}[2]{{\textstyle\frac{#1}{#2}}}
\begin{document} \title{Soft interactions in Herwig++}
\author{Manuel B\"ahr$^1$\protect\footnote{\ \ 
Talk given at First International Workshop on Multiple Partonic Interactions
at the LHC, ``MPI@LHC'08'', Perugia, Italy, October 27--31 2008},
 Jonathan M. Butterworth$^2$,
 Stefan Gieseke$^1$, Michael H. Seymour$^{3,4}$}
\institute{$^1$Institut f\"ur Theoretische Physik, Universit\"at Karlsruhe,\\ 
$^2$Department of Physics and Astronomy, University College London,\\ 
$^3$School of Physics and Astronomy, University of Manchester,\\ 
$^4$Physics Department, CERN.}
\maketitle
\begin{abstract}
We describe the recent developments to extend the multi-parton
interaction model of underlying events in \hwpp\ into the soft,
non-perturbative, regime.  This allows the program to describe also
minimum bias collisions in which there is no hard interaction, for the
first time.  It is publicly available from versions 2.3 onwards and
describes the Tevatron underlying event and minimum bias data.  The
extrapolations to the LHC nevertheless suffer considerable ambiguity, as
we discuss.
\end{abstract}

\section{Introduction}
\graphicspath{{./pics/hardUE/}} 

In this talk, we will summarize the development of a new model for the
underlying event in \hwpp, extending the previous perturbative
multi-parton interaction (MPI) model down into the soft non-perturbative
region.  This allows minimum bias collisions to be simulated by
\hwpp\ for the first time.

We begin, though, by mentioning a few of the features that accompanied
it in the release of \hwpp\cite{Bahr:2008pv} version
2.3\cite{Bahr:2008tf} in December 2008, which include NLO corrections in
the POWHEG scheme for single W and Z production\cite{Hamilton:2008pd},
and Higgs production\cite{Hamilton:2009za}.  Lepton--hadron scattering
processes have been included for the first time.  The simulation of 
physics beyond the standard model (BSM) has been extended to include a
much wider range of 3-body decays
and off-shell effects\cite{Gigg:2008yc}.  The treatment of baryon decays
has been extended to match the sophistication of meson and tau decays,
including off-shell and form factor effects and spin correlations.
Finally, in addition to the soft interactions discussed here, the MPI
model has been extended to include the possibility of selecting
additional scatters of arbitrary type, which can be important
backgrounds to BSM signatures for which the single-scattering
backgrounds are small, for example two like-sign Drell-Yan W
productions\cite{ManuelThesis}.

The semi-hard MPI model was implemented in \hwpp\ version
2.1\cite{Bahr:2007ni}.  It allows for the simulation of underlying
events with perturbative scatters with $p_t>\ptmin$ according to the
standard QCD matrix elements with standard PDFs, dressed by parton
showers that, in the initial state, account for the modifications of the
proton structure due to momentum and flavour conservation.  It
essentially re-implemented the existing Jimmy
algorithm\cite{Butterworth:1996zw} that worked with the fortran HERWIG
generator\cite{Corcella:2000bw}, but gave a significantly better
description of the CDF data on the underlying
event\cite{Affolder:2001xt,*Acosta:2004wqa}, in part due to a more
detailed global tuning\cite{Bahr:2008dy}.  However it was only able to
describe the jet production part of the data, above about 20~GeV, and
not the minimum bias part, owing to a lack of soft scatters below
\ptmin.  A possible extension into the soft regime was first discussed
in Ref.~\cite{Borozan:2002fk}, but we have provided the first robust
implementation of it, described in detail in Ref.~\cite{ManuelThesis}.
It is somewhat complementary to the approach used in
Pythia\cite{Sjostrand:1987su,Sjostrand:2004pf}, where the perturbative
scatters are extended into the soft region through the use of a smooth
non-perturbative modification.  However, we make a stronger connection
with information on total and elastic scattering cross sections,
available through the eikonal formalism, to place constraints on our
non-perturbative parameters\cite{Bahr:2008wk}.

\begin{wrapfigure}{r}{0.5\columnwidth}
  \vspace*{-0.5cm}
  \centerline{ \includegraphics[width=0.47\columnwidth]{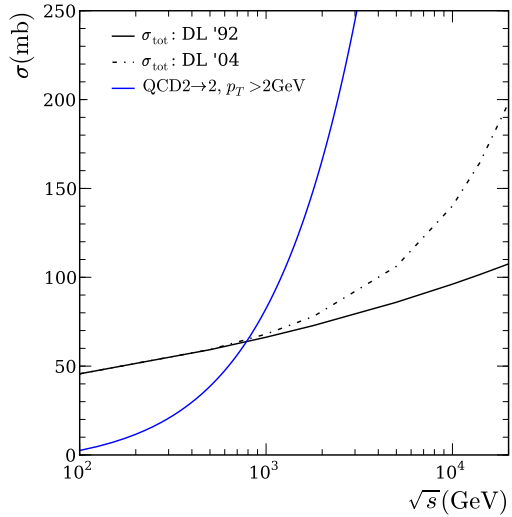} }
  \caption[]{Total cross sections (black) in the two parameterizations
  of Donnachie and Landshoff\cite{Donnachie:1992ny, Donnachie:2004pi}.
  In blue the QCD jet production cross section above 2 GeV is shown.}
  \label{fig:xsecs}
\end{wrapfigure}

In the remainder of this introduction, we recap the basics of the
eikonal model and recall the results of the perturbative MPI model that
we had previously implemented in \hwpp, before showing how to extend it
into the soft region.  In Sect.~\ref{sec:analytical} we discuss the
constraints that can be placed on the model by the connection with
hadronic scattering, and in Sect.~\ref{sec:finalstates} we show the
predictions for final state properties.

The starting point for the MPI model is the observation that the
inclusive cross section for perturbative parton scattering may exceed
the total hadron--hadron cross section.  We show an example in
Fig.~\ref{fig:xsecs}, with two of the total cross section
parameterizations we will be using.  The origin of the steep rise in the
partonic cross section is the proliferation of partons expected at small
$x$.  The excess of the partonic scattering cross section over the total
cross section simply implies that there is on average more than one
parton scattering per inelastic hadronic collision, $\bar
n=\sjet/\sinel$.  Since the majority of scatters come from very small
$x$ partons, they consume relatively little energy and it is a good
approximation to treat them as quasi-independent.

From the optical theorem, one derives a relationship between the Fourier
transform of the elastic amplitude $a(\vect{b},s)$ and the inelastic
cross section via the \emph{eikonal function,} $\chi(\vect{b},s)$,
\begin{equation}
  a(\vect{b},s)\equiv\frac1{2i}\left[
    \exp{-\chi(\vect{b},s)}-1\right]
  \qquad\longrightarrow\qquad
  \sinel = \int\di{\vect{b}}2\left[
    1-\exp{-2\chi(\vect{b},s)}\right].
\end{equation}
One can construct a QCD prediction for the eikonal function by assuming
that multiple scatters are independent, and that the partons that
participate in them are distributed across the face of the hadron with
some impact parameter distribution $G(\vect{b})$ that is independent of
their longitudinal momentum,
\begin{equation}
  \chi_{\rm QCD}(\vect{b},s)=\smfrac12\,A(\vect{b})\,\sigmahard(s),
  \qquad\qquad
  A(\vect{b}) =
  \int\di{\vect{b}'}2\,G(\vect{b}')\,G(\vect{b}-\vect{b}'),
\end{equation}
where \sigmahard\ is the inclusive partonic scattering cross section,
which is given by the conventional perturbative calculation.

\begin{figure}[tb]
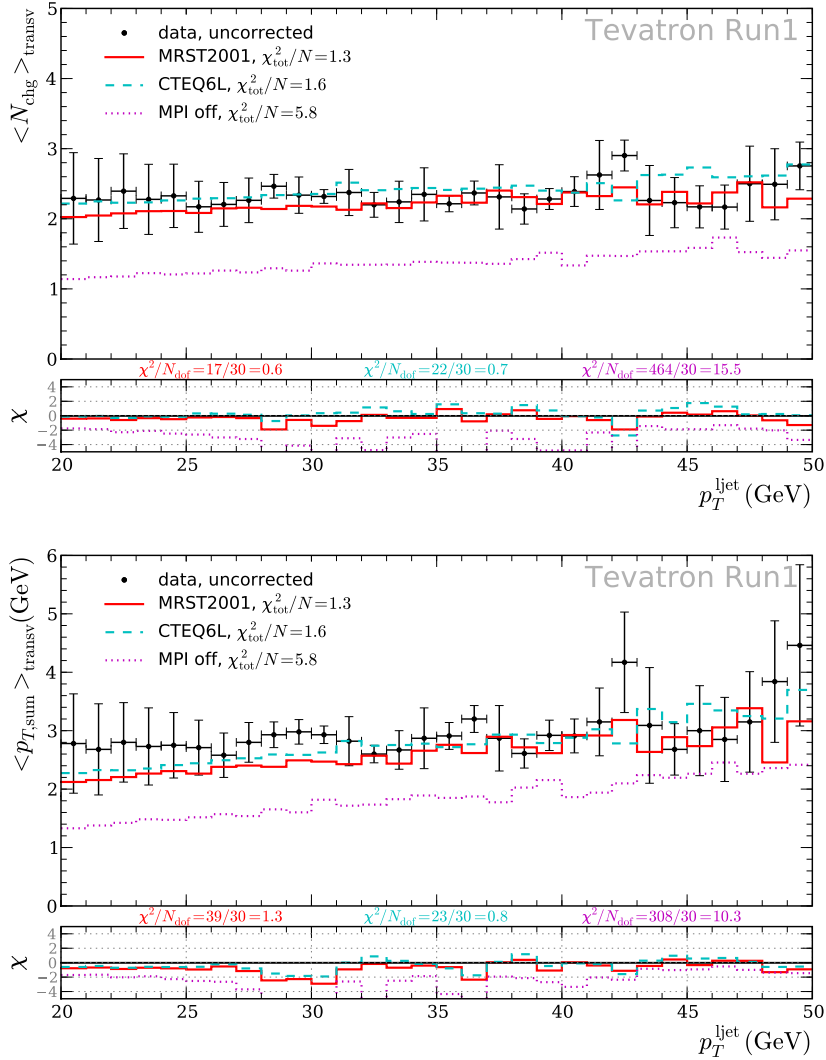

  \begin{center}
    \includegraphics[%
      width=.75\columnwidth,keepaspectratio]{nch_transv}
    \\[0.4cm]
    \includegraphics[%
      width=.75\columnwidth,keepaspectratio]{ptsum_transv}
  \end{center}
  \caption[Multiplicity and \ptsum\ in the transverse region.]{
    \label{fig:transv}
    Multiplicity and \ptsum\ in the \textbf{transverse} region. CDF data
    are shown as black circles, \hwpp\ without MPI as magenta dots, with
    MPI using MRST \cite{Martin:2001es} PDFs as solid red and with
    CTEQ6L \cite{Pumplin:2002vw} as cyan dashed. The lower plot shows
    the statistical significance of the disagreement between the Monte
    Carlo predictions and the data. The legend on the upper plot shows
    the total $\chi^2$ for all observables, whereas the lower plot for
    each observable has its $\chi^2$~values.}
\end{figure}

In the original Jimmy model and its \hwpp\ reimplementation, these
formulae are implemented in a straightforward way, with the hard cross
section defined by a strict cut, $p_t>\ptmin$ and the matter
distribution given by the Fourier transform of the electromagnetic form
factor,
\begin{equation}
  G(\vect{b}) = \int \frac{\di{\vect{k}}2}{(2\pi)^2}\;
  \frac{\exp{i\,\vect{k}\cdot\vect{b}}}{(1+\vect{k}^2/\musq)^2},
\end{equation}
with, to reflect the fact that the distribution of soft partons might
not be the same as that of electromagnetic charge, \musq\ considered to
be a free parameter and not fixed to its electromagnetic value
$0.71\GeV^2$.  Compared to a Gaussian of the same width, this
distribution has both a stronger peak and a broader tail so it is
somewhat similar to the double-Gaussian form used in
Pythia\cite{Sjostrand:2006za}.  In Ref.~\cite{Bahr:2008wk}, we
explicitly showed that the two result in similar distributions, if their
widths are fixed to be equal, except very far out in the tails.  \musq\
and \ptmin\ are the main adjustable parameters of the model and,
allowing them to vary freely, one can get a good description of the CDF
underlying event data, as shown in Fig.~\ref{fig:transv}.  The choice of
parton distribution function can also be seen to have a small but
significant effect.

The main shortcoming of this model is that it does not contain soft scatters
and hence cannot describe very low $p_t$ jet production or minimum bias
collisions.  In Ref.~\cite{Borozan:2002fk} it was proposed to remedy
this, by extending the concept of independent partonic scatters right
down into the infrared region.  One can therefore write the eikonal
function as the incoherent sum of the QCD component we already computed
and a soft component,
\begin{equation}
  \chi_{\rm tot}(\vect{b},s)=\chi_{\rm QCD}(\vect{b},s)+\chi_{\rm soft}(\vect{b},s)
  =\smfrac12\Bigl(A(\vect{b})\,\sigmahard(s)
  +A_{\rm soft}(\vect{b})\,\sigmasoft(s)\Bigr),
\end{equation}
where \sigmasoft\ is an unknown partonic soft scattering cross
section.  As a first simplest model, we assume that the matter
distributions are the same, $A_{\rm soft}(\vect{b})=A(\vect{b})$,
although we relax this condition later.  By taking the eikonal approach
seriously, we can trade the unknown soft cross section for the unknown
total hadronic cross section,
\begin{equation}
  \stot(s)=2\int\di{\vect{b}}2\left[
    1-\exp{-\frac12A(\vect{b})(\sigmahard(s)+\sigmasoft(s))}\right].
\end{equation}
Knowing the total cross section, for a given matter distribution and
hard cross section (implied by \ptmin\ and the PDF choice) the soft
cross section is then determined.  In order to make predictions for
energies higher than the Tevatron, we consider three predictions of the
total cross section: 1) the standard Donnachie--Landshoff
parameterization\cite{Donnachie:1992ny}; 2) the latter for the energy
dependence but with the normalization fixed by the CDF
measurement\cite{sigma_tot_CDF}; and 3) the newer Donnachie--Landshoff
model with a hard component\cite{Donnachie:2004pi}.  Of course once we
have an experimental measurement from the LHC we would use that for our
predictions.  In this way, our simple hard+soft model has no more free
parameters than our hard model and we can tune \musq\ and \ptmin.
Before doing this, we present the results of Ref.~\cite{Bahr:2008wk}, in
which we considered the theoretical constraints that could be put on
these parameters.

\section{Analytical constraints}
\label{sec:analytical}

\subsection{Simple model}
\graphicspath{{./pics/constraints/}} 

\begin{figure}[tb]
  \includegraphics[width=0.48\columnwidth]{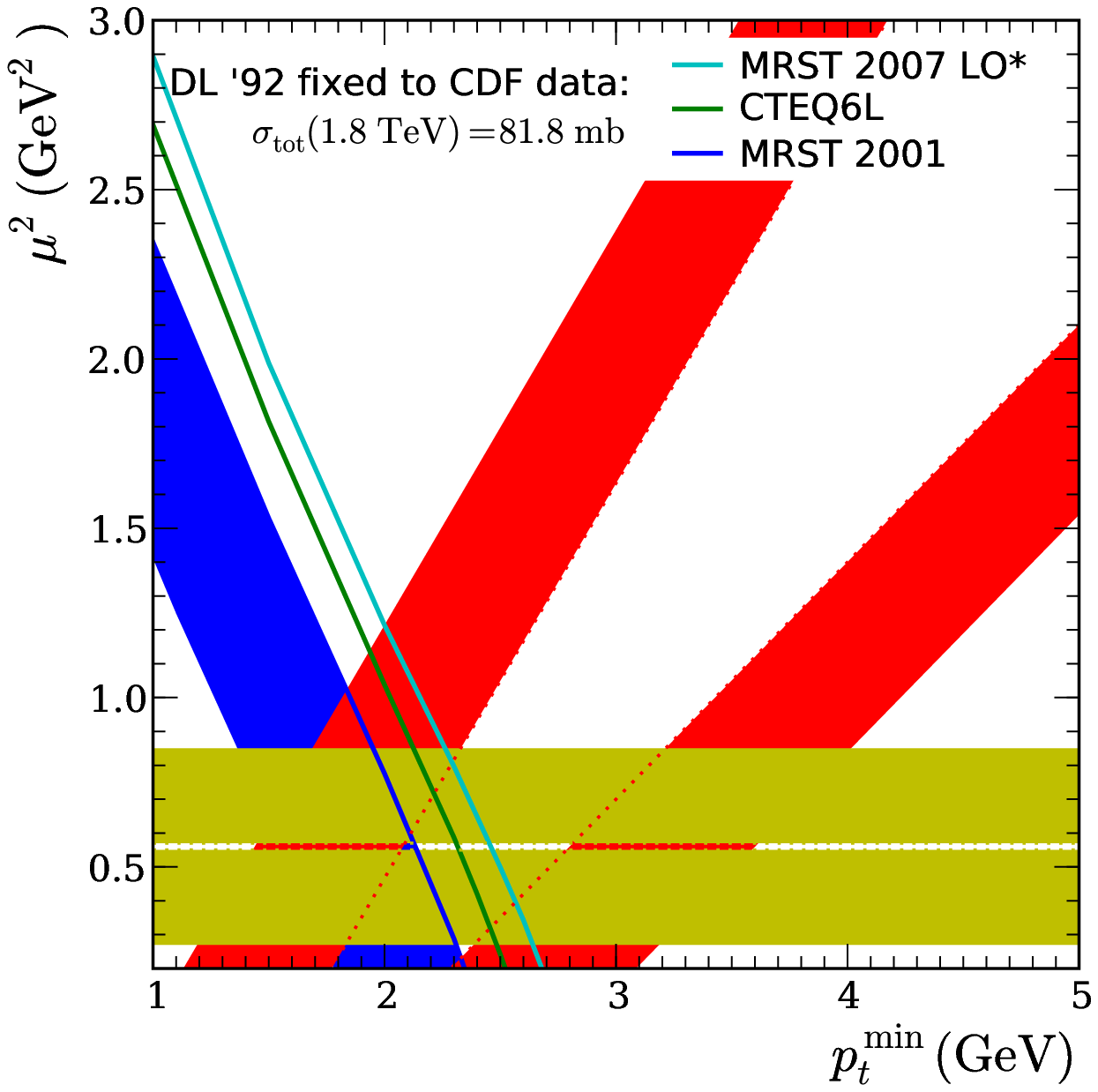}
  \includegraphics[width=0.48\columnwidth]{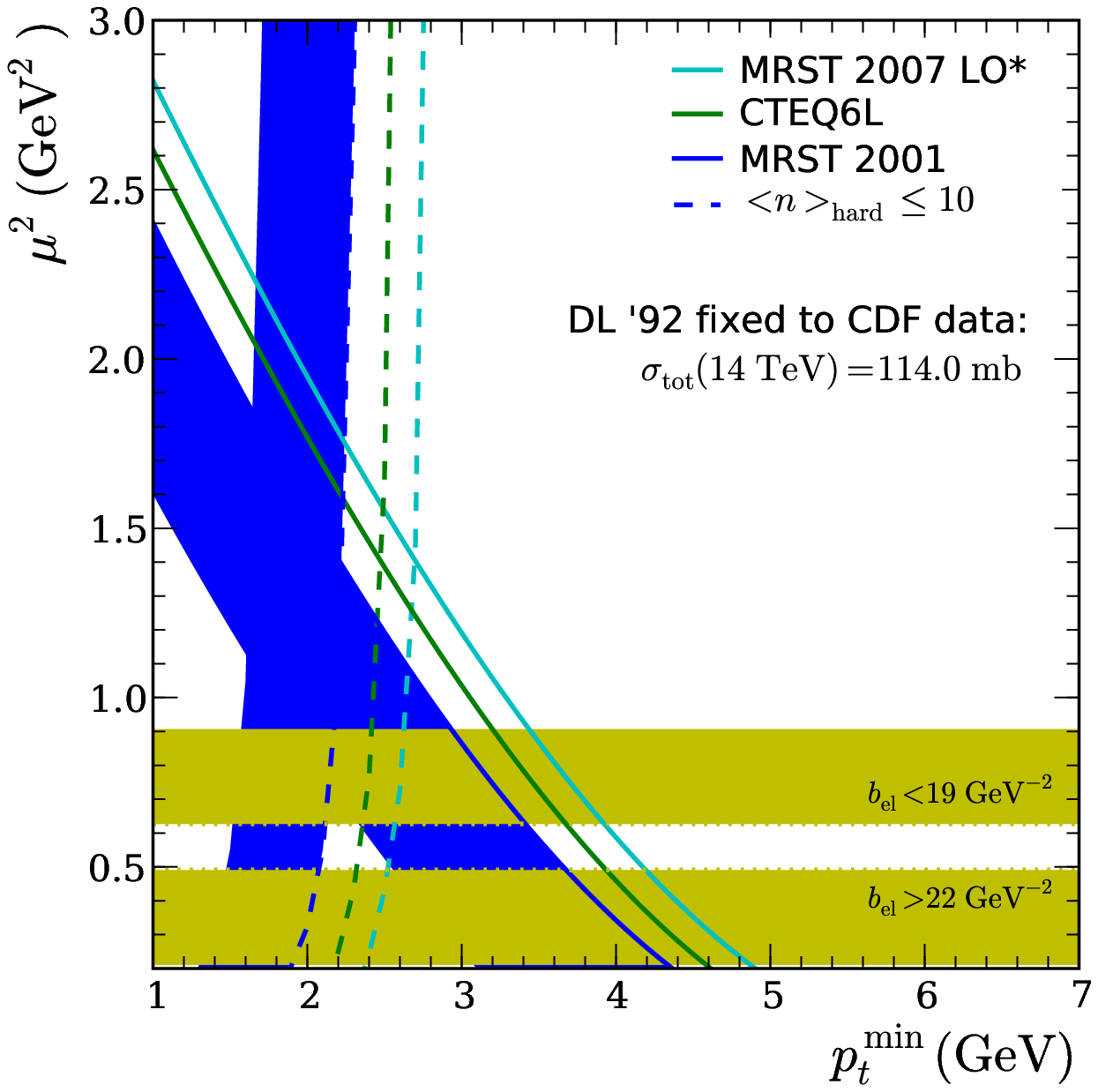}
  \caption[Parameter space at TVT]{Left: The parameter space of the
    simple eikonal model at the Tevatron. The solid curves come from
    $\sigmasoft>0$ for three different PDF sets. The horizontal lines
    come from $\slope = 16.98 \pm 0.25
    \GeV^{-2}$\cite{sigma_tot_CDF,Abe:1993xx}. The excluded regions are
    shaded. The dashed lines indicate the preferred parameter ranges
    from the fit to Tevatron final-state data\cite{Bahr:2008dy}.  Right:
    The equivalent plot for the LHC.  The additional (dashed)
    constraints come from requiring the total number of scatters to be
    less than 10.}
  \label{fig:constraints}
\end{figure}

Within our model we want \sigmasoft\ to correspond to a physical cross
section.  It must therefore be positive.  This therefore places
constraints on the \musq--\ptmin\ plane: a lower bound on \ptmin\ for
a given value of \musq.  These are shown for the Tevatron on the
left-hand side of Fig.~\ref{fig:constraints} as the solid lines for
three different PDF sets: the two shown previously and MRST
LO*\cite{Sherstnev:2007nd}.  Since in the eikonal model the total and
inelastic cross sections are related to the elastic one, we can also
place constraints from the elastic slope parameter, which has been
measured by CDF\cite{sigma_tot_CDF,Abe:1993xx}:
\begin{equation}
  \slope(s)\equiv\left[\frac{\rm d}{\di{t}{}}\left(\ln
      \frac{\rm d\sela}{\di{t}{}}\right)\right]_{t=0}
  =\frac1{\stot}\int\di{\vect{b}}2\,b^2
  \left[1-\exp{-\chi_{\rm tot}(\vect{b},s)}\right]
  =(17\pm0.25)\GeV^{-2}.
\end{equation}
This rather precise measurement directly constrains \musq\ in our
simple model and rules out all but a very narrow strip of the parameter
space.  Finally, we consider the parameter space of the fit to
final-state data.  Although there is a preferred point in the parameter
space, the tuning of both the hard-only model\cite{Bahr:2008dy} and the
hard+soft model shown below indicates a strong correlation between the
two parameters and there is a broad region of acceptable parameter
values, which we show in Fig.~\ref{fig:constraints} by the region edged
by red bands.  Between the different constraints we have only a very
small allowed region of parameter space.

At the LHC the picture is similar, although the constraint
$\sigmasoft>0$ is considerably more restrictive (note the difference in
range of the $x$~axes of the two plots).  Different models predict
$\slope$ in the range 19~to~22$\GeV^{-2}$ translating into a slightly
wider horizontal band.  Finally, although we do not have final-state
data to compare to, in order to simulate self-consistent final states at
all we find that we must prevent the multiplicity of scatters becoming
too high.  While precisely where we place this cut is arbitrary, we
indicate it by shading the region in which the mean number of scatters
is greater than~10.  This plot is shown for the central of the three
total LHC cross section predictions we consider~-- it is qualitatively
similar for the other two, although the different constraints move
somewhat.

Comparing the two plots in Fig.~\ref{fig:constraints}, we come to the
realization that, from these theoretical constraints together with the
fit to the Tevatron data, we can already rule out the possibility that
the parameters of this simple model are energy-independent~-- there is
no region of the plot that is allowed at both energies.

While it could be that the parameters of the MPI model are in fact
energy dependent, as advocated by the PYTHIA
authors\cite{Dischler:2000pk}, we prefer to let the LHC data decide, by
proposing a model that is flexible enough to allow energy-independent or
-dependent parameters.  The simplest generalization of the above model
that achieves this is actually well physically motivated, and we call it
the hot-spot model.

\subsection{Hot-Spot model}
\graphicspath{{./pics/softPlusHardUE/}} 

The simple model has other shortcomings, beyond our aesthetic preference
to allow the possibility of energy-independent parameters.  The values
of \sigmasoft\ extracted from the predictions of
$\stot$\cite{Bahr:2008wk}, have rather strange energy dependence, being
quite sensitive to precise details of the matter distribution, parameter
choice, cross section prediction and PDF set and, in most cases, having
a steeply rising dependence on energy, much steeper than one would like
to imagine for a purely soft cross section.  Moreover, the value of
\musq\ extracted from $\slope$ is in contradiction with that extracted
from CDF's measurement\cite{Abe:1997xk,*Treleani:2007gi,*comment} of
double-parton scattering, which yields $\musq=3.0\pm0.5\GeV^2$.

All of these shortcomings can be circumvented by allowing the matter
distribution to be different for soft and hard scatters.  As a next
simplest model, we keep the same form for each, but allow the \musq\
values to be different.  We again fix the additional free parameter,
this time to a fixed value of \slope.  That is, once $\stot$ and
$\slope$ are measured at some energy, the non-perturbative parameters of
our model, \sigmasoft\ and $\musq_{\rm soft}$ are known.  Since it will
turn out that our preferred value of \musq\ is significantly larger
than the extracted value of $\musq_{\rm soft}$, we call this a hot-spot
model: soft partons have a relatively broad distribution, actually
similar to the electromagnetic form factor, while semi-hard partons
(typically still small $x$, but probed at momentum scales above \ptmin)
are concentrated into smaller denser regions within the proton.

\begin{figure}[htb]
  \includegraphics[%
    width=0.48\textwidth]{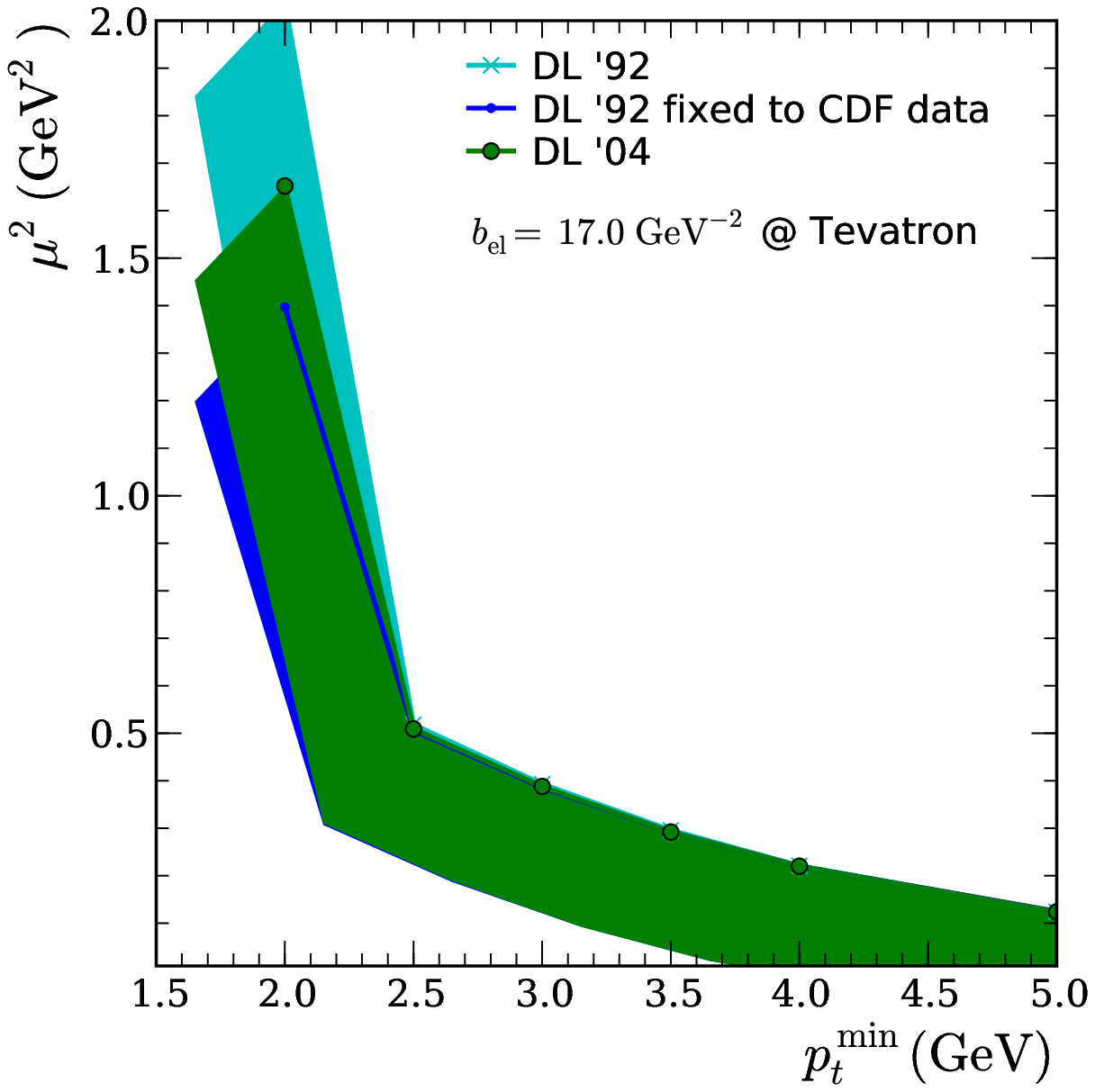}
  \includegraphics[%
    width=0.48\textwidth]{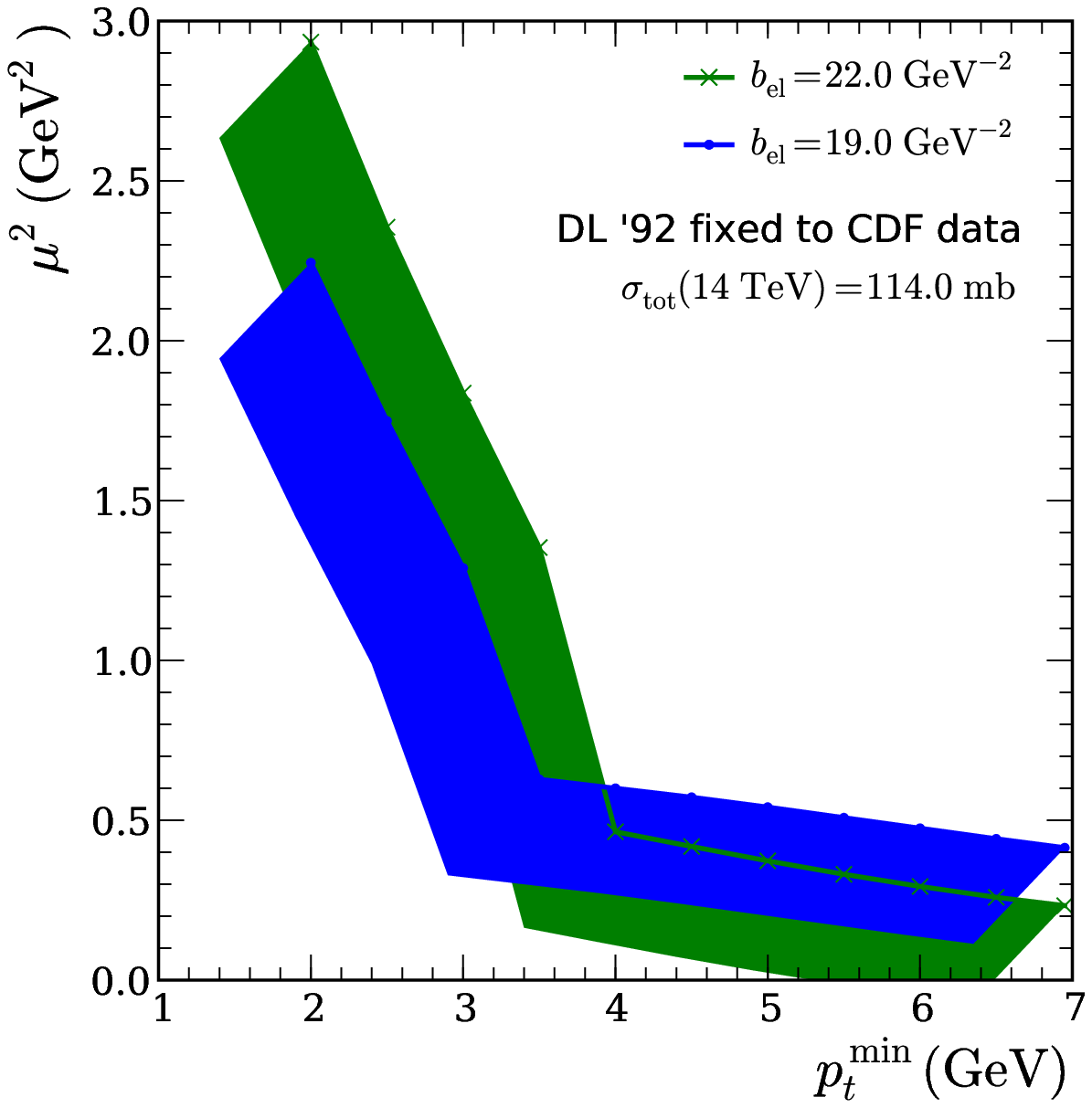}
  \caption[]{%
    \label{fig:constraints_imp}
    Parameter space of the improved eikonal model for the Tevatron
    (left) and LHC (right). The solid curves impose a minimum allowed
    value of \musq, for a given value of \ptmatch\ by requiring a valid
    description of \stot\ and \slope\ with positive \sigmasoft. The
    excluded regions are shaded. We used the MRST~2001
    LO\cite{Martin:2001es} PDFs for these plots.}
\end{figure}

Having used one constraint to fix an additional parameter, there is only
one constraint in the parameter space, shown in
Fig.~\ref{fig:constraints_imp} for the Tevatron and LHC.
The model has much more freedom than the simple one, with much of the
parameter space allowed, and with ample overlap between the allowed
regions at the two energies.

Another nice feature of this model is the energy-dependence of
\sigmasoft\ it implies, shown in Fig.~\ref{fig:sigma_soft_imp}.  At
least for the standard Donnachie--Landshoff energy dependence, it
corresponds to a very slow increase, almost constant, in-keeping with
one's expectations of a soft cross section.

\section{Final states}
\label{sec:finalstates}

We have implemented this model into \hwpp.  There are many additional
details that we do not go into here\cite{ManuelThesis}, but wherever
possible, the treatment of soft scatters is kept as similar as possible
to that of semi-hard scatters, to make for a smooth matching.  In
particular, for the transverse momentum dependence, we make the
distribution of $p_t^2$ a Gaussian centred on zero, whose integral over
the range zero to \ptmin\ is given by \sigmasoft\ and whose width is
adjusted such that $\di\sigma{}/\di{p_t}{}$ is continuous at \ptmin.
\ptmin\ is therefore seen to be not a cutoff, as it is in the Jimmy
model, but a matching scale, where the model makes a relatively smooth
transition between perturbative and non-perturbative treatments of the
same phenomena, in a similar spirit to the model of
Ref.~\cite{Gieseke:2007ad} for transverse momentum in initial-state
radiation.

The model actually exhibits a curious feature in its $p_t$ dependence,
first observed in Ref.~\cite{Borozan:2002fk}.  With the typical
parameter values that are preferred by the data,
$\di\sigma{}/\di{p_t}{}$ is large enough, and \sigmasoft\ small enough,
that the soft distribution is not actually a Gaussian but an inverted
Gaussian: its width-squared parameter is negative.  The result is that
the transverse momentum of scatters is dominated by the region around
\ptmin\ and not by the truly non-perturbative region $p_t\to0$.  This
adds to the self-consistency of the model, justifying the use of an
independent partonic scattering picture even for soft non-perturbative
collisions.

\begin{figure}[htb]
  \begin{minipage}[t]{0.48\textwidth}
  \includegraphics[width=\textwidth]{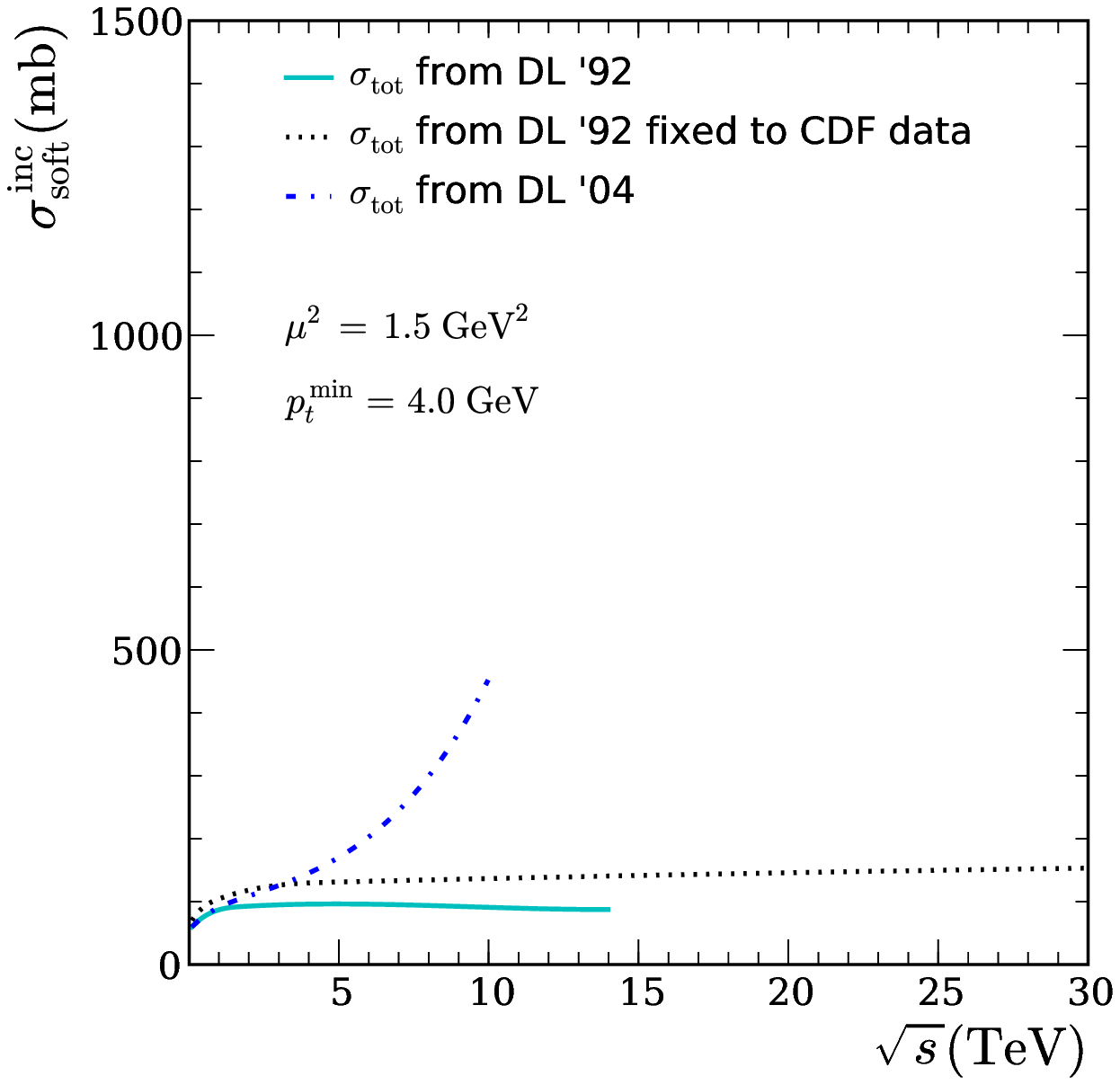}
  \caption[The extracted soft cross section in the improved model as a
    function of the centre-of-mass energy.]{
    \label{fig:sigma_soft_imp}
    \sigmasoft\ as a function of energy. Each of the three different
    curves shows the soft cross section that would appear when the
    respective parameterization for the total cross section is used.
    Curves that do not reach out to 30 TeV correspond to parameter
    choices that are unable to reproduce \stot\ and \slope\ correctly at
    these energies.}
  \end{minipage}
\graphicspath{{./pics/softPlusHardUE/}} 
  \begin{minipage}[t]{0.48\textwidth}
  \includegraphics[width=\textwidth,keepaspectratio]{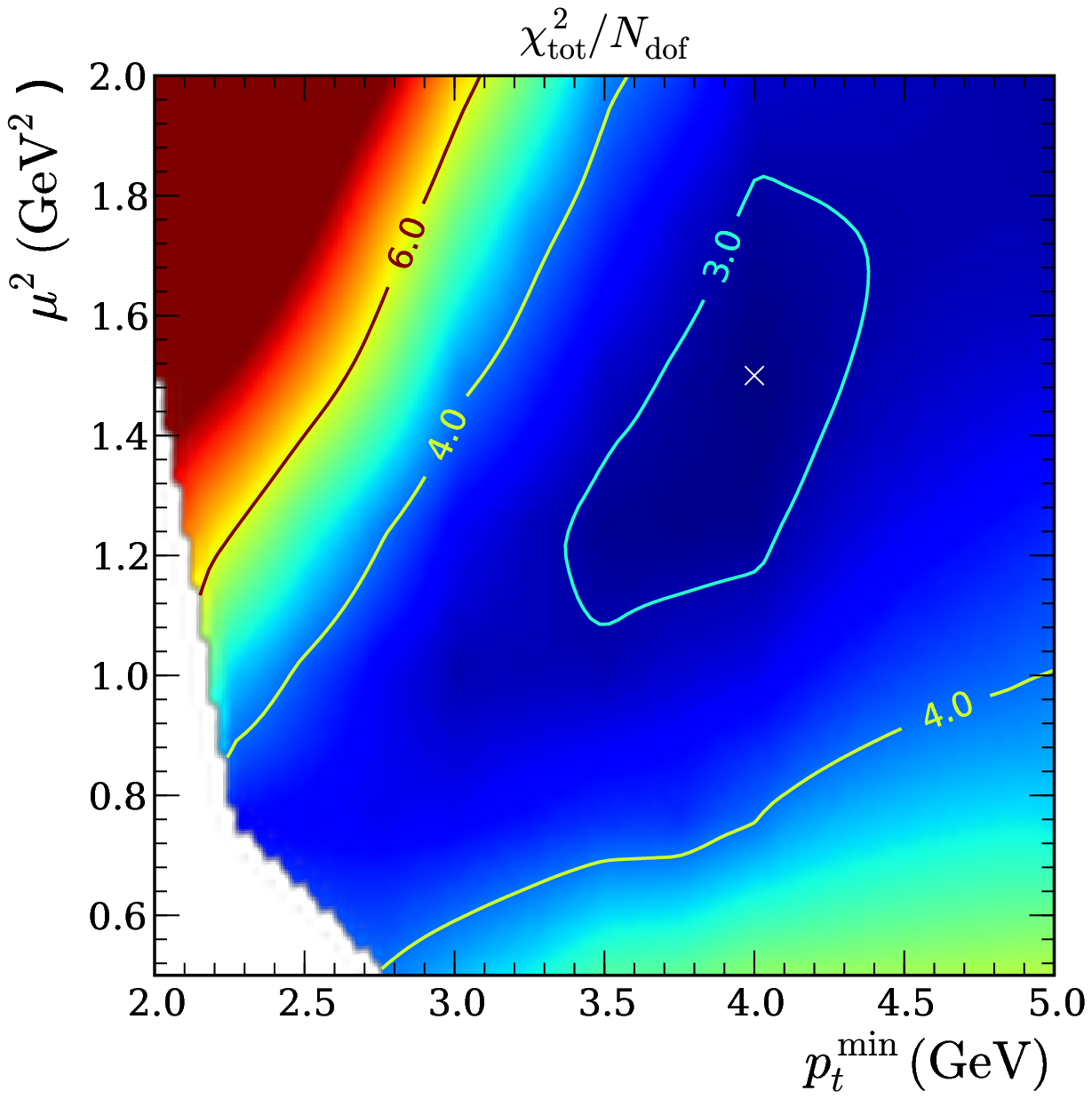}
  \caption[$\chi^2$-contour plots for all available bins]{
    \label{fig:scans_all}
    Contour plots for the $\chi^2$ per degree of freedom for the fit to
    the CDF underlying event data. The cross indicates the location of
    our preferred tune and the white area consists of parameter choices
    where the elastic $t$-slope and the total cross section cannot be
    reproduced simultaneously.}
  \end{minipage}
\end{figure}

\begin{figure}[tb]
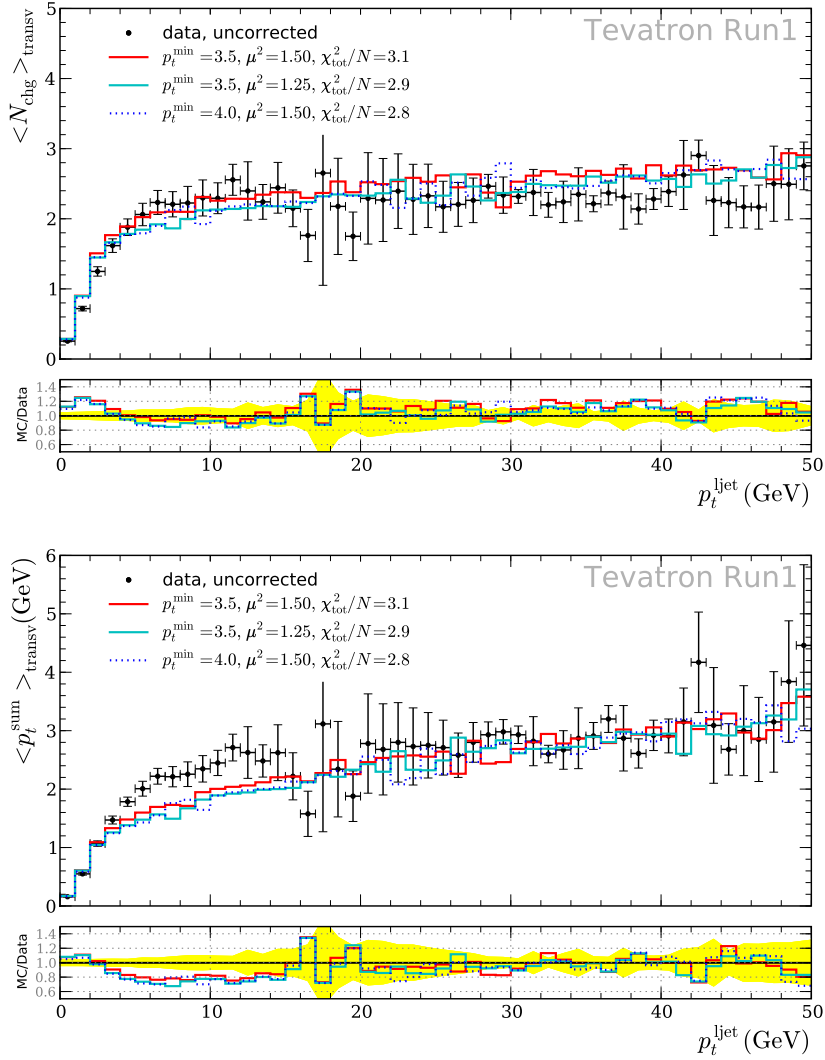

  \begin{center}
    \includegraphics[%
      width=.75\columnwidth,keepaspectratio]{nch_transv}
    \\[0.4cm]
    \includegraphics[%
      width=.75\columnwidth,keepaspectratio]{ptsum_transv}
  \end{center}
  \caption[Multiplicity and \ptsum\ in the transverse region.]{
    \label{fig:transv_soft}
    Multiplicity and \ptsum\ in the \textbf{transverse} region. CDF data
    are shown as black circles. The histograms show \hwpp\ with the
    improved model for semi-hard and soft additional scatters using the MRST
    2001 LO \cite{Martin:2001es} PDFs for three different parameter
    sets. The lower plot shows the ratio Monte Carlo to data and the
    data error band. The legend shows the total
    $\chi^2$ for all observables.}
\end{figure}

With the model in hand, we can repeat the tune to the CDF data on the
underlying event.  Unlike with the semi-hard model, we now fit the data
right down to zero leading jet momentum.  The result is shown in
Fig.~\ref{fig:scans_all}, which is qualitatively similar to the one for
the semi-hard model.
The description of the data in the transverse region is shown
in Fig.~\ref{fig:transv_soft}.
It can be seen to be reasonable in the lower transverse momentum region,
although certainly still not as good as at higher transverse momenta.

The discrepancy in the lowest few bins may be related to another
deficiency of our model.  According to the eikonal model, the inelastic
cross section should include all final states that are not exactly
elastic, while our simulation of them generates only non-diffractive
events in which colour is exchanged between the two protons and hence a
significant number of final-state hadrons are produced.  While
single-diffractive-dissociation events would not be triggered on
experimentally, double-diffractive-dissociation events, in which both
protons break up but do not exchange colour across the central region of
the event, would, and would lead to extremely quiet events with low
leading jet $p_t$ and low central multiplicity, which are not present in
our sample.  In Ref.~\cite{ManuelThesis} we have checked that these bins
are not pulling our tune significantly by repeating it without them.
The overall chi-squared is significantly smaller, but the best fit point
and chi-squared contours are similar.

\section{Conclusions}

We have reviewed the basis of the semi-hard MPI model that we previously
implemented in \hwpp, and motivated its extension to a soft component.
Through the connection with the total and elastic cross sections
provided by the eikonal model and optical theorem, we have placed
significant constraints on the simplest soft model.  We have shown that
these constraints can be relaxed by invoking a hot-spot model in which
the spatial distributions of soft and semi-hard partons are different.
Finally, we have implemented this model and shown that it gives a
reasonable description of the minimum bias data, for the first time in
\hwpp.  Nevertheless, there is still room for improvement, particularly
in the very low $p_t$ region and several avenues for further study
present themselves, not least the diffractive component already
mentioned, and the role of colour correlations, which were argued to be
very important in Ref.~\cite{Sjostrand:2004pf}, but which seem to be
less so in the current \hwpp\ implementation\cite{ManuelThesis}.

Despite the successful description of Tevatron data, the extrapolation
to the LHC suffers from considerable uncertainty.  The unknown value of
the total cross section, which determines the non-perturbative
parameters in our model, plays a crucial role, but even once this and
the elastic slope parameter have been directly measured, the region of
allowed parameter space is still large.  Although we prefer a model in
which the parameters are energy independent, ultimately only data will
tell us whether this is the case.  Finally, even once the underlying
event data have been measured, the parameters will not be fully tied
down, due to their entanglement with the PDFs.  We eagerly await the LHC
data to guide us.

\section*{Acknowledgements}

We are grateful to the other \hwpp\ authors and Leif L\"onnblad for
extensive collaboration.  MB and JMB gratefully acknowledge the
organizers of the First Workshop on Multiple Parton Interactions at the
LHC for a stimulating and productive meeting.  This work was supported
in part by the European Union Marie Curie Research Training Network
\emph{MCnet} under contract MRTN-CT-2006-035606 and the
Helmholtz Alliance ``Physics at the Terascale''.

\begin{footnotesize}
\bibliographystyle{mpi08} 
{\raggedright
\bibliography{perugia}
}
\end{footnotesize}
\end{document}